\documentclass[%
 reprint,
 amsmath,amssymb,
 aps,
pra,
]{revtex4-2}

\usepackage{graphicx}
\usepackage{dcolumn}
\usepackage{bm}

\begin{document}

\preprint{APS/123-QED}

\title{Towards dense optical communications with partially coherent light: \\II. Multimode Stokes communications}

\author{Ayman F. Abouraddy}
\email{raddy@creol.ucf.edu}
\author{Bahaa E. A. Saleh}%
\affiliation{CREOL, The College of Optics \& Photonics, University of Central Florida, Orlando, Florida 32816, USA}%

\date{\today}

\begin{abstract}
Partially coherent light has a higher informational-carrying capacity than coherent light when both fields are spanned by the same modes, which implies that partially coherent light may increase the channel density in optical communications. A partially coherent field spanned by $N_{\mathrm{f}}$ modes can multiplex $N_{\mathrm{D}}=N_{\mathrm{f}}^{2}-1$ data streams in the generalized Stokes parameters that uniquely identify its coherence matrix -- a scheme we call `Stokes communications' (S-Comm). We have proposed in an accompanying paper a realization of S-Comm for bimodal fields $N_{\mathrm{f}}=2$ that multiplex $N_{\mathrm{D}}=3$ data streams. Here we extend this proposed strategy for S-Comm to multimode fields $N_{\mathrm{f}}>2$. We find that the diagonal and off-diagonal Stokes parameters require different treatments and we provide detailed designs for the transmitter and receiver for both diagonally and off-diagonally multiplexed data streams. The building blocks of the proposed system can be readily implemented in integrated photonic circuit platforms. Our results indicate that on-chip processing of multimode partially coherent light (or `structured coherence') may have profound implications for optical communications and for optical signal processing in general.
\end{abstract}

\maketitle


\section{Introduction}

All the physical degrees of freedom (DoFs) of the optical field have been exploited for multiplexing in optical communications; e.g., the polarization DoF in polarization multiplexing \cite{Ip07JLT}, the spectral DoF in wavelength division multiplexing (WDM) \cite{Brackett90IEEEJSAC,Agrawal21Book}, and the spatial DoF in spatial-mode multiplexing \cite{Winzer13chapter,Richardson13NP,Li14AOP,Willner15AOP}. In each case, the DoF is spanned by a finite number of orthogonal modes, each of which is modulated by a data stream. Because mode orthogonality is critical to enable the unique separation (demultiplexing) of combined (multiplexed) modulated modes without crosstalk, a DoF spanned by $N_{\mathrm{f}}$~modes multiplexes $\sim N_{\mathrm{f}}$ data steams.

The last DoF of the optical field that has yet to be exploited for multiplexing is optical coherence. A partially coherent optical field spanned by $N_{\mathrm{f}}$ modes is described by an $N_{\mathrm{f}}\times N_{\mathrm{f}}$ Hermitian coherence matrix $\mathbf{G}$ that requires $N_{\mathrm{D}}=N_{\mathrm{f}}^{2}-1$ real parameters for its unique identification (with one parameter reserved for normalization). One possible parameterization of such a coherence matrix makes use of the generalized Stokes parameters (SPs), which are $N_{\mathrm{f}}^{2}$ real parameters (including normalization) that uniquely identify the coherence matrix. Modulating each SP with a data stream multiplexes $\mathcal{O}(N_{\mathrm{f}}^{2})$ data streams in a field supported by only $N_{\mathrm{f}}$~modes, thus promising a dramatic increase in the channel capacity. We call such a scheme `Stokes communications' (S-Comm). We described in Part~I of this work \cite{Abouraddy26JOSAA1} a detailed construction of an S-Comm system that makes use of bimodal light $N_{\mathrm{f}}=2$ to encode $N_{\mathrm{D}}=N_{\mathrm{f}}^{2}-1=3$ data streams. The system described in \cite{Abouraddy26JOSAA1} makes use of the conventional definition of SPs to parameterize the $2\times2$ coherence matrix.

Here we extend S-Comm to multimode $N_{\mathrm{f}}>2$ partially coherent light to encode $\mathcal{O}(N_{\mathrm{f}}^{2})$ data streams. To this end, we first introduce generalized SPs that are the real expansion coefficients of the $N_{\mathrm{f}}\times N_{\mathrm{f}}$ coherence matrix $\mathbf{G}$ in terms of generalized Pauli matrices (also known as generalized Gell-Mann matrices) \cite{Kimura03PLA,Bertlmann08JPA}. A crucial distinction emerges between the generalized Pauli matrices associated with the $N_{\mathrm{S}}=N_{\mathrm{f}}(N_{\mathrm{f}}-1)$ off-diagonal elements of $\mathbf{G}$ and those associated with its $N_{\mathrm{f}}$ diagonal elements. The former (off-diagonal) Pauli matrices correspond to the conventional off-diagonal $2\times2$ Pauli matrices after embedding them in the $N_{\mathrm{f}}\times N_{\mathrm{f}}$ space. Consequently, multiplexing data streams in the off-diagonal elements of $\mathbf{G}$ makes use of the same bimodal system in \cite{Abouraddy26JOSAA1} as a building block. In contrast, the standard definition of generalized Pauli matrices associated with the diagonal SPs do not map to the diagonal $2\times2$ Pauli matrix. Consequently, multiplexing data streams in the diagonal SPs requires developing new building blocks. Besides introducing optical realizations of these new building blocks, we offer two alternatives. In one alternative, we retain as much as possible the conventional $2\times2$ Pauli matrix structure in the $N_{\mathrm{f}}\times N_{\mathrm{f}}$ setting. In a second alternative, we introduce a hybrid strategy, whereby the off-diagonal SPs make use of generalized Pauli matrices for multiplexing $N_{\mathrm{S}}=N_{\mathrm{f}}(N_{\mathrm{f}}-1)$ data streams, while the diagonal elements are modulated directly with $N_{\mathrm{f}}$ data streams via on-off keying as is typical in optical multiplexing without recourse to diagonal SPs. The hybrid approach is the simplest in terms of the construction of the multiplexer and demultiplexer. However, a drawback is that the power varies in the optical signal from bit to bit, whereas the power level is constant when relying on the generalized SPs.

Crucially, the building blocks of the proposed multimodal S-Comm system can be implemented with standard components in integrated photonic circuits. This is a marked departure from the focus on freely propagating fields that has been the hallmark of the study of partial coherence since its inception \cite{Zernicke,Born99Book,Wolf07Book,Gbur10PO,Korotkova20PO}. This has changed over the past year with the emergence of experimental reports of on-chip manipulation of partially coherent multimode light; see Refs.~\cite{Hashemi26APLPhotonTwoMode,Hashemi26arxiv4Modes,Hashemi26LPR3modes,Hashemi26OL,Mor26arxiv}. On-chip implementations provide the necessary interferometric stability and small footprint that guarantee the scalability of S-Comm. The results outlined here lay the foundation for a new form of optical communications that makes use of partially coherent light and promises profound advances in optical information processing.

\section{Conventional bimodal Stokes parameters}\label{sec:BimodalSPs}

Bimodal ($N_{\mathrm{f}}=2$) partially coherent optical fields are described by a $2\times2$ Hermitian, positive semi-definite, unity-trace coherence matrix $\mathbf{G}$ written in a basis of modes $|1\rangle$ and $|2\rangle$ (in the Dirac notation) as $\mathbf{G}=\left(\!\begin{array}{cc}G_{11}&G_{12}\\G_{21}&G_{22}\end{array}\!\right)$, where $G_{jk}=\langle E_{j}E_{k}^{*}\rangle=G_{kj}^{*}$, $j,k=1,2$, $E_{j}$ is the complex amplitude associated with the mode $|j\rangle$, and $\langle\cdot\rangle$ represents an ensemble average. The eigenvalues $\lambda_{1}$ and $\lambda_{2}$ of $\mathbf{G}$ are real, $\lambda_{1},\lambda_{2}\geq0$, and $\mathrm{Tr}\{\mathbf{G}\}=G_{11}+G_{22}=\lambda_{1}+\lambda_{2}=1$. The diagonal elements $G_{11}$ and $G_{22}$ are the fractions of power associated with modes $|1\rangle$ and $|2\rangle$, whereas the off-diagonal element $G_{12}$ represents their statistical correlation. Detectors that project onto the modes register the modal weights $G_{11}$ and $G_{22}$, while interferometry is needed to reveal $G_{12}$ (see accompanying paper \cite{Abouraddy26JOSAA1}).

A useful parameterization of $\mathbf{G}$ is in terms of the SPs, which are the real expansion coefficients of $\mathbf{G}$ in terms of Pauli matrices:
\begin{equation}\label{eq:BimodalSPs}
\mathbf{G}=\frac{1}{2}\sum_{j=0}^{3}s_{j}\hat{\sigma}_{j}=\frac{1}{2}\left(\begin{array}{cc}s_{0}+s_{3}&s_{1}-is_{2}\\s_{1}+is_{2}&s_{0}-s_{3}\end{array}\right),
\end{equation}
where $\{\hat{\sigma}_{j}\}$ are the Pauli matrices:
\begin{equation}
\hat{\sigma}_{1}=\!\left(\!\begin{array}{cc}0&1\\1&0\end{array}\!\right)\!,\;
\hat{\sigma}_{2}=\!\left(\!\begin{array}{cc}0&-i\\i&0\end{array}\!\right)\!,\;
\hat{\sigma}_{3}=\!\left(\!\begin{array}{cc}1&0\\0&-1\end{array}\!\right)\!,\;
\end{equation}
and $\hat{\sigma}_{0}=\hat{\mathbb{I}}_{2}$ is the $2\times2$ identity matrix. We define the degree of coherence $D=\lambda_{1}-\lambda_{2}$ of $\mathbf{G}$, with $0\leq D\leq1$ when taking $1\geq\lambda_{1}\geq\lambda_{2}\geq0$.

The key properties of the Pauli matrices for our purposes here are: (1) $\mathrm{Tr}\{\hat{\sigma}_{j}\}=0$ ($j\neq0$); and (2) $\mathrm{Tr}\{\hat{\sigma}_{j}\hat{\sigma}_{k}\}=2\delta_{jk}$ ($j,k=1,2,3$). These two properties enable us to extract the SPs from $\mathbf{G}$ through the projection $s_{j}=\mathrm{Tr}\{\hat{\sigma}_{j}\mathbf{G}\}$, so that $s_{1}=2\mathrm{Re}\{G_{12}\}$, $s_{2}=-2\mathrm{Im}\{G_{12}\}$, $s_{3}=G_{11}-G_{22}$, and $s_{0}=G_{11}+G_{22}=1$; conversely, $G_{11}=1+s_{3}$, $G_{22}=1-s_{3}$, and $G_{12}=s_{1}-is_{2}$. In general, $|s_{j}|\leq1$ and $0\leq s_{1}^{2}+s_{2}^{2}+s_{3}^{2}=D^{2}\leq1$ \cite{Abouraddy26AOP}.

\subsection{Bimodal Stokes multiplexing}

\begin{figure}[t!]
\centering
\includegraphics[width=8.8cm]{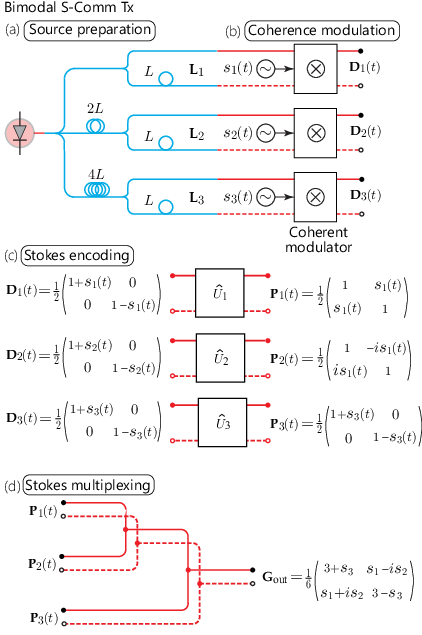}
\caption{Bimodal S-Comm transmitter. (a) Source preparation. Three identically, mutually uncorrelated bimodes $\mathbf{L}_{j}=\tfrac{1}{2}\hat{\mathbb{I}}_{2}$ are prepared from an optical source of finite coherence length, $j=1,2,3$. (b) Coherence modulation of $\mathbf{L}_{j}$ via the data streams $s_{j}(t)$ to yield coherence-modulated bimodes $\mathbf{D}_{j}(t)$. (c) Stokes-encoding unitaries $\hat{U}_{j}$ convert $\mathbf{D}_{j}(t)$ into Stokes-encoded bimodes $\mathbf{P}_{j}(t)=\hat{U}_{j}\mathbf{D}_{j}(t)\hat{U}_{j}^{\dagger}$. The diagonal elements in $\mathbf{D}_{j}(t)$ are `moved' to a prescribed position in $\mathbf{P}_{j}(t)$. (d) Stokes multiplexing. The first modes underpinning $\mathbf{P}_{1}(t)$, $\mathbf{P}_{2}(t)$, and $\mathbf{P}_{3}(t)$ are combined into $|1\rangle$, and the second modes into $|2\rangle$.}
\label{Fig:BimodalTX}
\end{figure}

The bimodal field $\mathbf{G}$ is uniquely identified by three real parameters (its SPs), each of which can be exploited as an information carrier in the dense optical communications system we call S-Comm, where $N_{\mathrm{D}}=N_{\mathrm{f}}^{2}-1$ data streams are multiplexed in only $N_{\mathrm{f}}$~modes. For a bimodal field, $N_{\mathrm{D}}=3$ data streams are multiplexed in $N_{\mathrm{f}}=2$ modes. We depict in Fig.~\ref{Fig:BimodalTX} the major blocks of the bimodal S-Comm transmitter. First, three mutually uncorrelated bimodes are prepared by splitting an optical source of linewidth $\Delta\lambda$ and coherence length $L_{\mathrm{o}}$ into 6~paths and inserting relative optical lengths such that any two paths differ by at least $L>L_{\mathrm{o}}$. One example is depicted in Fig.~\ref{Fig:BimodalTX}(a), whereupon the relative lengths inserted into the 6~paths are $0,L,2L,3L,4L$, and $5L$; see Ref.~\cite{Abouraddy26JOSAA1} for details. Each bimode is described by the same coherence matrix $\mathbf{L}=\tfrac{1}{2}\hat{\mathbb{I}}_{2}$, that is modulated by a data stream $s_{j}(t)$ to yield coherence-modulated bimodes $\mathbf{D}_{j}(t)$, $j=1,2,3$ [Fig.~\ref{Fig:BimodalTX}(b)]:
\begin{equation}
\mathbf{D}_{j}(t)=\frac{1}{2}\left(\begin{array}{cc}1+s_{j}(t)&0\\0&1-s_{j}(t)\end{array}\right).
\end{equation}
The binary data streams $s_{j}(t)$ can take on the values~0 and~1 (conventional on-off keying) or~1 and~$-1$. This procedure is realized by a `coherence modulator' whose embodiments are described in Ref.~\cite{Abouraddy26JOSAA1}. Stokes-encoding unitaries $\hat{U}_{j}$:
\begin{equation}
\hat{U}_{1}=\frac{1}{\sqrt{2}}\!\left(\!\begin{array}{cc}1&-1\\1&1\end{array}\!\right)\!,\;
\hat{U}_{2}=\frac{1}{\sqrt{2}}\!\left(\!\begin{array}{cc}1&i\\i&1\end{array}\!\right)\!,\;
\hat{U}_{3}=\hat{\mathbb{I}}_{2},
\end{equation}
convert the coherence-modulated bimodes $\mathbf{D}_{j}(t)$ into Stokes-encoded bimodes $\mathbf{P}_{j}(t)=\hat{U}_{j}\mathbf{D}_{j}(t)\hat{U}_{j}^{\dagger}$:
\begin{eqnarray}
\mathbf{P}_{1}(t)\!\!\!&=&\!\!\!\frac{1}{2}\left(\!\!\begin{array}{cc}1&s_{1}(t)\\s_{1}(t)&1\end{array}\!\!\right)\!,\;
\mathbf{P}_{2}(t)=\frac{1}{2}\left(\!\!\begin{array}{cc}1&-is_{2}(t)\\is_{2}(t)&1\end{array}\!\!\right)\!,\nonumber\\
\mathbf{P}_{3}(t)\!\!\!&=&\!\!\!\frac{1}{2}\left(\!\!\begin{array}{cc}1+s_{3}(t)&0\\0&1-s_{j}(t)\end{array}\!\!\right).
\end{eqnarray}
Consequently, $s_{1}(t)$ and $s_{2}(t)$ are now associated with the off-diagonal elements of the bimodes, whereas $s_{3}(t)$ remains along the diagonal [Fig.~\ref{Fig:BimodalTX}(c)]. Finally, the pair of modes underpinning each of the three bimodes are combined to form a bimodal field $\mathbf{G}_{\mathrm{out}}(t)=\tfrac{1}{3}\{\mathbf{P}_{1}(t)+\mathbf{P}_{2}(t)+\mathbf{P}_{3}(t)\}$:
\begin{equation}
\mathbf{G}_{\mathrm{out}}(t)=\frac{1}{6}\left(\begin{array}{cc}
3+s_{3}(t)&s_{1}(t)-is_{2}(t)\\s_{1}(t)+is_{2}(t)&3-s_{3}(t)
\end{array}\right).
\end{equation}
We refer to this step as `Stokes multiplexing' [Fig.~\ref{Fig:BimodalTX}(d)]. The bimodal field $\mathbf{G}_{\mathrm{out}}(t)$ is then sent over the optical communications channel. The modes $|1\rangle$ and $|2\rangle$ may correspond to two single-mode fibers, two guided modes in a multimode fiber, two orthogonal spatial modes in a free-space or underwater link, or two orthogonal polarization modes \cite{Abouraddy26JOSAA1}.

\subsection{Bimodal Stokes demultiplexing}

\begin{figure}[t!]
\centering
\includegraphics[width=7.5cm]{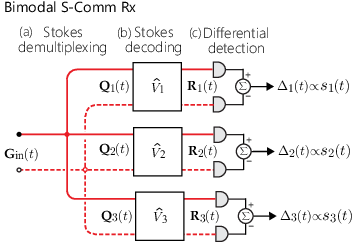}
\caption{Receiver for bimodal S-Comm. (a) Stokes demultiplexing. Mode $|1\rangle$ in $\mathbf{G}_{\mathrm{in}}(t)$ is split into three copies, and similarly for $|2\rangle$. One copy from both sets comprise the bimode $\mathbf{Q}_{j}(t)$, $j=1,2,3$. (b) Stokes-decoding unitaries $\hat{V}_{j}$ convert $\mathbf{Q}_{j}(t)$ into the bimode $\mathbf{R}_{j}(t)=\hat{V}_{j}\mathbf{Q}_{j}(t)\hat{V}_{j}^{\dagger}$. (c) Differential detection. The two modal weights (the diagonal elements) of $\mathbf{R}_{j}(t)$ are recorded, and their difference $\Delta_{j}(t)\propto s_{j}(t)$ are obtained.}
\label{Fig:BimodalRX}
\end{figure}

The bimodal S-Comm receiver, which reverses the operations in the bimodal S-Comm transmitter, is illustrated in Fig.~\ref{Fig:BimodalRX}. First, the received bimodal field $\mathbf{G}_{\mathrm{in}}(t)\approx\mathbf{G}_{\mathrm{out}}(t)$ is split into three identical bimodes $\mathbf{Q}_{j}(t)=\mathbf{G}_{\mathrm{in}}(t)$, $j=1,2,3$, a step we call `Stokes demultiplexing' [Fig.~\ref{Fig:BimodalRX}(a)]. Each bimode $\mathbf{Q}_{j}(t)$ is directed to a Stokes-decoding unitary $\hat{V}_{j}=\hat{U}_{j}^{\dagger}$ [Fig.~\ref{Fig:BimodalRX}(b)], to convert it into the bimode $\mathbf{R}_{j}(t)=\hat{V}_{j}\mathbf{Q}_{j}(t)\hat{V}_{j}^{\dagger}$, given by:
\begin{eqnarray}\label{eq:BimodalRmatrices}
\mathbf{R}_{1}(t)\!\!\!&=&\!\!\!\frac{1}{6}\left(\!\!\begin{array}{cc}3+s_{1}(t)&-s_{3}(t)-is_{2}(t)\\-s_{3}(t)+is_{2}(t)&3-s_{1}(t)\end{array}\!\!\right)\!,\nonumber\\
\mathbf{R}_{2}(t)\!\!\!&=&\!\!\!\frac{1}{6}\left(\!\!\begin{array}{cc}3+s_{2}(t)&s_{1}(t)+is_{3}(t)\\s_{1}(t)-is_{3}(t)&3-s_{2}(t)\end{array}\!\!\right)\!,\nonumber\\
\mathbf{R}_{3}(t)\!\!\!&=&\!\!\!\frac{1}{6}\!\left(\!\!\begin{array}{cc}3+s_{3}(t)&s_{1}(t)-is_{2}(t)\\s_{1}(t)+is_{2}(t)&3-s_{3}(t)\end{array}\!\!\right)\!.
\end{eqnarray}
Each bimode $\mathbf{R}_{j}(t)$ is directed to a differential detector [Fig.~\ref{Fig:BimodalRX}(c)] to obtain the difference $\Delta_{j}$ between its modal weights, whereby:
\begin{equation}
\Delta_{1}(t)\propto s_{1}(t),\;\Delta_{2}(t)\propto s_{2}(t),\;\Delta_{3}(t)\propto s_{3}(t),
\end{equation}
and the three data streams are retrieved without crosstalk, thereby completing the bimodal S-Comm protocol.


\section{Extending the Stokes parameters to multimode light}

Our goal here is to describe an extension for S-Comm to a multimode field $N_{\mathrm{f}}>2$, which can be mostly decomposed into the bimodal S-Comm building blocks from Fig.~\ref{Fig:BimodalTX} and Fig.~\ref{Fig:BimodalRX}. This requires first defining generalized multimode SPs.

A multimode ($N_{\mathrm{f}}>2$) partially coherent optical field spanned by the modes $|j\rangle$, $j=1,\cdots,N_{\mathrm{f}}$, is described by the $N_{\mathrm{f}}\times N_{\mathrm{f}}$ coherence matrix $\mathbf{G}$:
\begin{equation}\label{eq:NxNcoherenceMatrix}
\mathbf{G}=\left(\begin{array}{cccc}
G_{11}&G_{12}&\cdots&G_{1,N_{\mathrm{f}}}\\
G_{21}&G_{22}&\cdots&G_{2,N_{\mathrm{f}}}\\
\vdots&\vdots&\ddots&\vdots\\
G_{N_{\mathrm{f}},1}&G_{N_{\mathrm{f}},2}&\cdots&G_{N_{\mathrm{f}},N_{\mathrm{f}}}\\
\end{array}\right),
\end{equation}
where the definition of the elements $G_{jk}$ follows that of its $2\times2$ counterpart (except for the range of $j$ and $k$). Therefore, $\mathbf{G}$ is Hermitian, positive semi-definite, and unity-trace, so that many of the properties listed above follow; e.g., $\sum_{j=1}^{N_{\mathrm{f}}}G_{jj}=\sum_{j=1}^{N_{\mathrm{f}}}\lambda_{j}=1$, where $\{\lambda_{j}\}$ are the real eigenvalues of $\mathbf{G}$. The coherence matrix has $\tfrac{1}{2}N_{\mathrm{f}}(N_{\mathrm{f}}-1)$ complex off-diagonal elements in the upper triangle of $\mathbf{G}$, which require $N_{\mathrm{S}}=N_{\mathrm{f}}(N_{\mathrm{f}}-1)$ real parameters for their identification. Two parameters are needed per each of these off-diagonal elements corresponding to their real and imaginary parts. Additionally, $\mathbf{G}$ has $N_{\mathrm{f}}$ real diagonal elements. The total number of real parameters to identify $\mathbf{G}$ is thus $N_{\mathrm{S}}+N_{\mathrm{f}}=N_{\mathrm{f}}^{2}$, which is reduced to $N_{\mathrm{D}}=N_{\mathrm{f}}^{2}-1$ once the trace is normalized.

We extend the parameterization of a $2\times2$ coherence matrix $\mathbf{G}$ in terms of SPs (Eq.~\ref{eq:BimodalSPs}) to $N_{\mathrm{f}}>2$ as follows:
\begin{equation}
\mathbf{G}=\frac{1}{2}\sum_{j=0}^{N_{\mathrm{D}}}s_{j}\hat{\Lambda}_{j},
\end{equation}
where $\{s_{j}\}$ are the (real) generalized modal SPs and $\{\hat{\Lambda}_{j}\}$ are generalized Pauli matrices of dimensions $N_{\mathrm{f}}\times N_{\mathrm{f}}$, with $\hat{\Lambda}_{0}=\hat{\mathbb{I}}_{N_{\mathrm{f}}}$ the $N_{\mathrm{f}}\times N_{\mathrm{f}}$ identity matrix. The modal set $\{\hat{\Lambda}_{j}\}$ is a basis for the space of $N_{\mathrm{f}}\times N_{\mathrm{f}}$ matrices. When $\mathbf{G}$ is Hermitian, the expansion coefficients $\{s_{j}\}$ are real. We require that the modal set $\{\hat{\Lambda}_{j}\}$ has the same critical properties we listed above for the conventional Pauli matrices:
\begin{enumerate}
\item $\mathrm{Tr}\{\hat{\Lambda}_{j}\}=0$, $j\neq0$. Consequently, $\mathrm{Tr}\{\mathbf{G}\}=\tfrac{1}{2}s_{0}\mathrm{Tr}\{\hat{\mathbb{I}}_{N_{\mathrm{f}}}\}=\tfrac{1}{2}s_{0}N_{\mathrm{f}}=1$, so that $s_{0}=\tfrac{2}{N_{\mathrm{f}}}$.
\item $\mathrm{Tr}\{\hat{\Lambda}_{j}\hat{\Lambda}_{k}\}=2\delta_{jk}$, with $j,k=1,\cdots,N_{\mathrm{D}}$.
\end{enumerate}
These two properties allow us to extract the SPs from $\mathbf{G}$ though the same projection as above: $s_{j}=\mathrm{Tr}\{\hat{\Lambda}_{j}\mathbf{G}\}$.

To relate the generalized SPs $\{s_{j}\}$ to the elements of $\mathbf{G}$, we need to define the generalized Pauli matrices $\{\hat{\Lambda}_{j}\}$. We start by dividing the set $\{\hat{\Lambda}_{j}\}$ into two classes:
\begin{enumerate}
\item \textit{Off-diagonal matrices} that have zeros along the diagonal and are thus related only to the complex off-diagonal elements of $\mathbf{G}$.  We have $N_{\mathrm{S}}=N_{\mathrm{f}}(N_{\mathrm{f}}-1)$ such matrices.
\item \textit{Diagonal matrices} that have non-zero elements only along the diagonal (all off-diagonal elements are zero) and are thus related to the real diagonal elements of $\mathbf{G}$. We have $N_{\mathrm{f}}-1$ such matrices (excluding $\hat{\Lambda}_{0}$).
\end{enumerate}
We thus have $N_{\mathrm{S}}$ off-diagonal and $N_{\mathrm{f}}-1$ diagonal SPs, for a total of $N_{\mathrm{D}}$. We proceed to describe these two classes in detail.


\section{Off-diagonal generalized Pauli matrices}

These are $N_{\mathrm{f}}\times N_{\mathrm{f}}$ matrices with the diagonal elements set to zero. We model these matrices after the $\hat{\sigma}_{1}$ and $\hat{\sigma}_{2}$ Pauli matrices. We divide these $N_{\mathrm{S}}=N_{\mathrm{f}}(N_{\mathrm{f}}-1)$ off-diagonal matrices into $\tfrac{1}{2}N_{\mathrm{S}}$ pairs, one pair for each off-diagonal element \cite{Kimura03PLA,Bertlmann08JPA}. We index the SPs starting from the element $G_{12}$ in $\mathbf{G}$ (Eq.~\ref{eq:NxNcoherenceMatrix}). The real and imaginary parts of the the element $G_{12}$ corresponding to the statistical correlations between the modes $|1\rangle$ and $|2\rangle$ are associated with the SPs $s_{1}$ and $s_{2}$. We define this off-diagonal element as $G_{12}=s_{1}-is_{2}$, corresponding to the two generalized off-diagonal Pauli matrices $\hat{\Lambda}_{1}$ and $\hat{\Lambda}_{2}$:
\begin{equation}
\hat{\Lambda}_{1}=\left(\begin{array}{cccc}
0&1&0&\cdots\\
1&0&0&\cdots\\
0&0&0&\cdots\\
\vdots&\vdots&\vdots&\ddots
\end{array}\right)\!,\;
\hat{\Lambda}_{2}=\left(\begin{array}{cccc}
0&-i&0&\cdots\\
i&0&0&\cdots\\
0&0&0&\cdots\\
\vdots&\vdots&\vdots&
\end{array}\right)\!.
\end{equation}
The only non-zero elements of $\hat{\Lambda}_{1}$ and $\hat{\Lambda}_{2}$ are the ones displayed, so that $\mathrm{Tr}\{\hat{\Lambda}_{j}\hat{\Lambda}_{k}\}=0$. Next, the off-diagonal element $G_{13}=s_{3}-is_{4}$ in $\mathbf{G}$, which is associated with the matrices $\hat{\Lambda}_{3}$ and $\hat{\Lambda}_{4}$:
\begin{equation}
\hat{\Lambda}_{3}=\left(\begin{array}{cccc}
0&0&1&\cdots\\
0&0&0&\cdots\\
1&0&0&\cdots\\
\vdots&\vdots&\vdots&
\end{array}\right)\!,\;
\hat{\Lambda}_{4}=\left(\begin{array}{cccc}
0&0&-i&\cdots\\
0&0&0&\cdots\\
i&0&0&\cdots\\
\vdots&\vdots&\vdots&
\end{array}\right)\!.
\end{equation}
We continue to move along the off-diagonal elements of the first row of $\mathbf{G}$ ending at $G_{1,N_{\mathrm{f}}}=s_{2N_{\mathrm{f}}-3}-is_{2N_{\mathrm{f}}-2}$, associated with the two matrices:
\begin{equation}
\hat{\Lambda}_{2N_{\mathrm{f}}-3}=\!\left(\!\!\begin{array}{cccc}
0&\cdots&0&1\\
0&\cdots&0&0\\
\vdots&\vdots&\vdots&\vdots\\
1&\cdots&0&0
\end{array}\!\!\right)\!,\;
\hat{\Lambda}_{2N_{\mathrm{f}}-2}=\!\left(\!\!\begin{array}{cccc}
0&\cdots&0&-i\\
0&\cdots&0&0\\
\vdots&\vdots&\vdots&\vdots\\
i&\cdots&0&0
\end{array}\!\!\right)\!.
\end{equation}
We then move on to the off-diagonal element $G_{23}$ in the second row of $\mathbf{G}$, and so on to the last off-diagonal element $G_{N_{\mathrm{f}}-1,N_{\mathrm{f}}}=s_{N_{\mathrm{S}}-1}-is_{N_{\mathrm{S}}}$, which is associated with the two matrices:
\begin{equation}
\hat{\Lambda}_{N_{\mathrm{S}}-1}=\!\left(\!\!\begin{array}{cccc}
0&\cdots&0&0\\
\vdots&\vdots&\vdots&\vdots\\
0&\cdots&0&1\\
0&\cdots&1&0
\end{array}\!\!\right),\;
\hat{\Lambda}_{N_{\mathrm{S}}}=\!\left(\!\!\begin{array}{cccc}
0&\cdots&0&0\\
\vdots&\vdots&\vdots&\vdots\\
0&\cdots&0&-i\\
0&\cdots&i&0
\end{array}\!\!\right).
\end{equation}

In general, the off-diagonal element $G_{jk}$ ($j<k$; corresponding to the elements in the upper triangle of $\mathbf{G}$ above the diagonal) can be written as $G_{jk}=s_{2\ell-1}(t)-is_{2\ell}(t)$, where $j,k=1,\cdots,N_{\mathrm{f}}$, $\ell=1,\cdots,\tfrac{1}{2}N_{\mathrm{S}}$, and:
\begin{equation}\label{eq:ell}
\ell=(j-1)\left(N_{\mathrm{f}}-\frac{j}{2}\right)+k-j.
\end{equation}
Therefore, each off-diagonal element $G_{jk}$ is associated with the two generalized Pauli matrices $\hat{\Lambda}_{2\ell-1}$ and $\hat{\Lambda}_{2\ell}$.

Crucially, each off-diagonal element of $\mathbf{G}$ is associated with a pair of matrices that are basically the $2\times2$ Pauli matrices $\hat{\sigma}_{1}$ and $\hat{\sigma}_{2}$ embedded into the $N_{\mathrm{f}}\times N_{\mathrm{f}}$ space. This vastly simplifies the structure of the multimode S-Comm transmitter and receiver associated with off-diagonal SPs, which are constructed by utilizing the off-diagonal block in bimodal S-Comm as a unit cell for each pair of modes.

\subsection{Off-diagonal Stokes multiplexing}

The machinery introduced in Section~\ref{sec:BimodalSPs} for multiplexing the two data streams $s_{1}(t)$ and $s_{2}(t)$ into the off-diagonal element of a bimodal field ($N_{\mathrm{f}}=2$) can be repeated for \textit{each} off-diagonal element of an $N_{\mathrm{f}}\times N_{\mathrm{f}}$ coherence matrix for a multimode field $N_{\mathrm{f}}>2$. We follow here for the multimode S-Comm transmitter the same steps illustrated in Fig.~\ref{Fig:BimodalTX} for the bimodal S-Comm transmitter except for scaling up the number of initial bimodes to $N_{\mathrm{S}}$.

\begin{figure}[t!]
\centering
\includegraphics[width=8.8cm]{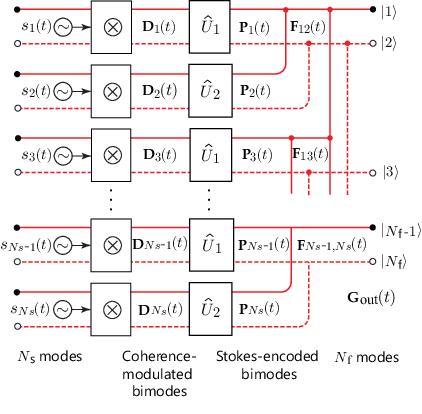}
\caption{Design for an S-Comm transmitter that multiplexes $N_{\mathrm{S}}=N_{\mathrm{f}}(N_{\mathrm{f}}-1)$ data streams in the off-diagonal SPs.}
\label{Fig:SCommOffDiagTX}
\end{figure}

We start by preparing $N_{\mathrm{S}}$ bimodes. Each of these initially identical but uncorrelated bimodes (all described by the coherence matrix $\mathbf{L}=\tfrac{1}{2}\hat{\mathbb{I}}_{2}$) is directed to a coherence modulator along with a data stream $s_{j}(t)$ to produce a coherence-modulated bimode $\mathbf{D}_{j}(t)$, $j=1,\cdots,N_{\mathrm{S}}$ [Fig.~\ref{Fig:SCommOffDiagTX}]. Each of the coherence-modulated bimodes $\mathbf{D}_{j}(t)$ is directed to a Stokes-encoding unitary, which alternate between $\hat{U}_{1}$ and $\hat{U}_{2}$, thus yielding $N_{\mathrm{S}}$ Stokes encoded bimodes $\mathbf{P}_{j}(t)$. Each pair of such bimodes (Stokes-encoded by $\hat{U}_{1}$ and $\hat{U}_{2}$) are then combined into one bimode that will contribute to a single off-diagonal element of the final coherence matrix $\mathbf{G}_{\mathrm{out}}(t)$. The bimodes $\mathbf{P}_{1}(t)$ and $\mathbf{P}_{2}(t)$ are combined into bimode $\mathbf{F}_{12}(t)$, and the bimodes $\mathbf{P}_{3}(t)$ and $\mathbf{P}_{4}(t)$ are combined into the bimode $\mathbf{F}_{13}(t)$, where:
\begin{eqnarray}
\mathbf{F}_{12}(t)&=&\frac{1}{4}\left(\begin{array}{cc}
2&s_{1}(t)-is_{2}(t)\\s_{1}(t)+is_{2}(t)&2
\end{array}\right),\nonumber\\
\mathbf{F}_{13}(t)&=&\frac{1}{4}\left(\begin{array}{cc}
2&s_{3}(t)-is_{4}(t)\\s_{3}(t)+is_{4}(t)&2
\end{array}\right),
\end{eqnarray}
and so on until all the bimodes $\mathbf{P}_{j}(t)$ are used, yielding $\tfrac{1}{2}N_{\mathrm{S}}$ bimodes of the form $\mathbf{F}_{jk}(t)$, $j,k=1,\cdots,N_{\mathrm{f}}$, $j<k$. The indices of the bimode $\mathbf{F}_{jk}(t)$ indicate the position of the off-diagonal element of $\mathbf{G}_{\mathrm{out}}(t)$ that the bimode will contribute to. The final step is that of Stokes multiplexing. The $\tfrac{1}{2}N_{\mathrm{S}}$ bimodes $\mathbf{F}_{jk}(t)$ comprise $N_{\mathrm{S}}=N_{\mathrm{f}}(N_{\mathrm{f}}-1)$ modes, which are collected into $N_{\mathrm{f}}$ sets of $N_{\mathrm{f}}-1$ modes each. The modes in each set are to be combined into a single mode, thus leaving us with a total of $N_{\mathrm{f}}$~modes constituting $\mathbf{G}_{\mathrm{out}}(t)$. This procedure collects all the modes with index~1 from the bimodes $\mathbf{F}_{jk}(t)$ to become the mode $|1\rangle$, all the modes with the index~2 are collected to yield the modes $|2\rangle$, and so on to the modes with the index $N_{\mathrm{f}}$ that are collected into the mode $|N_{\mathrm{f}}\rangle$ [Fig.~\ref{Fig:SCommOffDiagTX}]. 

\subsection{Off-diagonal Stokes demultiplexing}

\begin{figure}[t!]
\centering
\includegraphics[width=8.8cm]{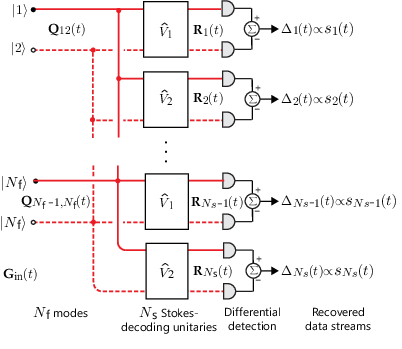}
\caption{Design for a multimode S-Comm receiver that demultiplexes $N_{\mathrm{S}}$ data streams from the off-diagonal SPs.}
\label{Fig:SCommOffDiagRX}
\end{figure}

At the multimode S-Comm receiver [Fig.~\ref{Fig:SCommOffDiagRX}], assuming $\mathbf{G}_{\mathrm{in}}(t)\approx\mathbf{G}_{\mathrm{out}}(t)$, all the steps in the multimode S-Comm transmitter are reversed. The first step is that of Stokes demultiplexing, where each received mode (from a total of $N_{\mathrm{f}}$ received modes) is divided into $N_{\mathrm{f}}-1$ copies, resulting in a total of $N_{\mathrm{S}}=N_{\mathrm{f}}(N_{\mathrm{f}}-1)$ modes. These modes are then combined into $\tfrac{1}{2}N_{\mathrm{S}}$ bimodes $\mathbf{Q}_{jk}(t)$ by combining pairs of modes obtained from $|j\rangle$ and $|k\rangle$. The off-diagonal element of each bimode $\mathbf{Q}_{jk}(t)$ contains the off-diagonal element $G_{jk}=s_{2\ell-1}(t)-is_{2\ell}(t)$ from $\mathbf{G}_{\mathrm{in}}(t)$, with $j<k$, $j,k=1,\cdots,N_{\mathrm{f}}$, $\ell=1,\cdots,\tfrac{1}{2}N_{\mathrm{S}}$, and $\ell$ is related to $j$ and $k$ through Eq.~\ref{eq:ell}.

In the next step, each bimode $\mathbf{Q}_{jk}(t)$ is split into two copies and directed to a pair of unitaries $\hat{V}_{1}=\hat{U}_{1}^{\dagger}$ and $\hat{V}_{2}=\hat{U}_{2}^{\dagger}$, which bring either the real part or the imaginary part of the off-diagonal element onto the diagonal. This results in $N_{\mathrm{S}}$ bimodes $\mathbf{R}_{j}(t)$, $j=1,\cdots,N_{\mathrm{S}}$, which are directed to differential detectors that record the difference $\Delta_{j}$ between the modal weights of each bimode $\mathbf{R}_{j}(t)$, yielding $\Delta_{j}(t)\propto s_{j}(t)$.

Three points must be emphasized here. (1) The multimode Stokes multiplexing and demultiplexing systems are built by repeating a unit cell comprising the off-diagonal bimodal system as a building block \cite{Abouraddy26JOSAA1}; see also Fig.~\ref{Fig:BimodalTX} and Fig.~\ref{Fig:BimodalRX}. Increasing the size of the system (increasing $N_{\mathrm{f}}$) requires simply adding more identical unit cells. Conversely, a subsection of a larger system can be utilized by not providing all the inputs. (2) All the components needed (Stokes-encoding unitaries, Stokes-decoding unitaries, Stokes multiplexing, Stokes demultiplexing) can be implemented on chip \cite{Capmany20Book,Bogaerts20Nature}. (3) The devices are all fixed; no changes are needed to be implemented from bit to bit or within a bit.


\begin{figure*}[t!]
\centering
\includegraphics[width=16.8cm]{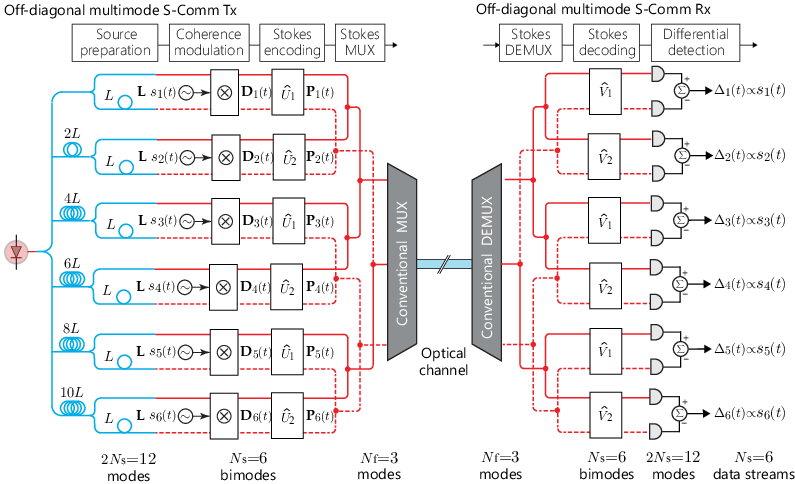}
\caption{Multimode S-Comm transmitter and receiver for $N_{\mathrm{f}}=3$ that utilize only the off-diagonal SPs. MUX: Multiplexer; DEMUX: demultiplexer.}
\label{Fig:3modeOffDiagTXRX}
\end{figure*}

\subsection{Example of three-mode light}

Consider the concrete example of $N_{\mathrm{f}}=3$ and $N_{\mathrm{S}}=6$. We thus have 6~Stokes-encoded bimodes $\mathbf{P}_{j}(t)$, which are collected in pairs: $\mathbf{P}_{1}(t)$ and $\mathbf{P}_{2}(t)$ are combined to yield a bimode $\mathbf{F}_{12}(t)$; $\mathbf{P}_{3}(t)$ and $\mathbf{P}_{4}(t)$ are combined to yield a bimode $\mathbf{F}_{13}(t)$; and $\mathbf{P}_{5}(t)$ and $\mathbf{P}_{6}(t)$ are combined to yield a bimode $\mathbf{F}_{23}(t)$. The three bimodes $\mathbf{F}_{12}(t)$, $\mathbf{F}_{13}(t)$, and $\mathbf{F}_{23}(t)$ each comprise a pair of modes. We combine modes $|1\rangle$ from $\mathbf{F}_{12}(t)$ and $\mathbf{F}_{13}(t)$, modes $|2\rangle$ from $\mathbf{F}_{12}(t)$ and $\mathbf{F}_{23}(t)$, and modes $|3\rangle$ from $\mathbf{F}_{13}(t)$ and $\mathbf{F}_{23}(t)$. This results in the three output modes $|1\rangle$, $|2\rangle$, and $|3\rangle$ that support the $3\times3$ field $\mathbf{G}_{\mathrm{out}}(t)$. Mathematically, the bimodes $\mathbf{F}_{12}(t)$, $\mathbf{F}_{13}(t)$, and $\mathbf{F}_{23}(t)$ are embedded into the space spanned by the modal basis $\{|1\rangle,|2\rangle,|3\rangle\}$, and are thus expressed as:
\begin{equation}
\mathbf{F}_{12}^{\mathrm{E}}(t)=\frac{1}{4}\left(\begin{array}{ccc}
2&s_{1}(t)-is_{2}(t)&0\\
s_{1}(t)+is_{2}&2&0\\
0&0&0
\end{array}\right),\nonumber
\end{equation}
\begin{equation}
\mathbf{F}_{13}^{\mathrm{E}}(t)=
\frac{1}{4}\left(\begin{array}{ccc}
2&0&s_{3}(t)-is_{4}(t)\\
0&0&0\\
s_{3}(t)+is_{4}(t)&0&2
\end{array}\right),\nonumber
\end{equation}
\begin{equation}
\mathrm{F}_{23}^{\mathrm{E}}(t)=
\frac{1}{4}\left(\begin{array}{ccc}
0&0&0\\
0&2&s_{5}(t)-is_{6}(t)\\
0&s_{5}(t)+is_{6}(t)&2
\end{array}\right).\nonumber
\end{equation}
Stokes-multiplexing combines the corresponding modes to yield the output field $\mathbf{G}_{\mathrm{out}}(t)=\tfrac{1}{3}\{\mathbf{F}_{12}^{\mathrm{E}}(t)+\mathbf{F}_{13}^{\mathrm{E}}(t)+\mathbf{F}_{23}^{\mathrm{E}}(t)\}$:
\begin{equation}
\mathbf{G}_{\mathrm{out}}(t)\!=\!\frac{1}{12}\!\left(\!\begin{array}{ccc}
4&s_{1}(t)-is_{2}(t)&s_{3}(t)-is_{4}(t)\\
s_{1}(t)+is_{2}(t)&4&s_{5}(t)-is_{6}(t)\\
s_{3}(t)+is_{4}(t)&s_{5}(t)+is_{6}(t)&4
\end{array}\!\right).
\end{equation}
Although $\mathbf{G}_{\mathrm{out}}(t)$ is spanned by $N_{\mathrm{f}}=3$ modes, it nevertheless multiplexes $N_{\mathrm{S}}=N_{\mathrm{f}}(N_{\mathrm{f}}-1)=6$ independent data streams in its off-diagonal elements. Note that the diagonal elements of $\mathbf{G}_{\mathrm{out}}(t)$ are all constant, so that modal detectors do not recover or detect any of these data streams, thereby yielding a secure form of communications as described in \cite{Abouraddy26JOSAA1} for bimodal fields (see also Ref.~\cite{Nardi22OL}).

Once the three modes supporting $\mathbf{G}_{\mathrm{in}}(t)$ are received, each mode is split into two copies, thus yielding 6~modes to be arranged into three bimodes: $\mathbf{Q}_{12}(t)$ that comprises a copy of $|1\rangle$ and $|2\rangle$; $\mathbf{Q}_{13}(t)$ that comprises a copy of $|1\rangle$ and $|3\rangle$; and $\mathbf{Q}_{23}(t)$ that comprises a copy of $|2\rangle$ and $|3\rangle$. The bimode $\mathbf{Q}_{12}(t)$ is divided among Stokes-decoding unitaries $\hat{V}_{1}$ and $\hat{V}_{2}$ to yield bimodes $\mathbf{R}_{1}(t)$ and $\mathbf{R}_{2}(t)$; $\mathbf{Q}_{13}(t)$ yields $\mathbf{R}_{3}(t)$ and $\mathbf{R}_{4}(t)$; and $\mathbf{Q}_{23}(t)$ yields $\mathbf{R}_{5}(t)$ and $\mathbf{R}_{6}(t)$. Each bimode $\mathbf{R}_{j}(t)$, $j=1,\cdots,6$, is directed to a differential detector that obtains the difference between the two modal weights, thus yielding $\Delta_{j}(t)\propto s_{j}(t)$.

This overall S-Comm scheme considering only the off-diagonal SPs when $N_{\mathrm{f}}=3$ is depicted in Fig.~\ref{Fig:3modeOffDiagTXRX}.


\section{Diagonal generalized Pauli matrices}

\begin{figure}[t!]
\centering
\includegraphics[width=8.8cm]{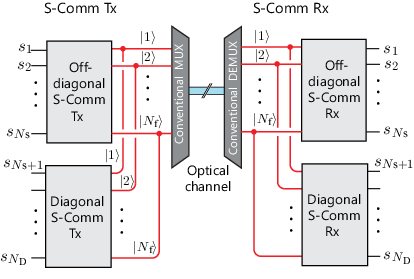}
\caption{Combining the $N_{\mathrm{f}}$-mode fields produced by the off-diagonal and diagonal S-Comm transmitters (depicted on the left). On the right, we depict the splitting of the received $N_{\mathrm{f}}$-mode field into two copies that are directed to the off-diagonal and diagonal S-Comm receivers.}
\label{Fig:CombiningDiafOffDiag}
\end{figure}

The off-diagonal S-Comm transmitter [Fig.~\ref{Fig:SCommOffDiagTX}] produces $N_{\mathrm{f}}$~modes carrying $N_{\mathrm{S}}$~data streams that are multiplexed into the optical channel. We discuss here the \textit{diagonal} S-Comm transmitter, which also produces a field spanned by $N_{\mathrm{f}}$ modes at its output and carries the remaining $N_{\mathrm{D}}-N_{\mathrm{S}}=N_{\mathrm{f}}-1$ data streams. In a multimode S-Comm transmitter that encodes data streams in the diagonal \textit{and} off-diagonal elements of $\mathbf{G}$, the outputs from these two sub-systems are combined into a single set of $N_{\mathrm{f}}$~modes, which are then multiplexed into the optical channel; see Fig.~\ref{Fig:CombiningDiafOffDiag}. The reverse is implemented at the S-Comm receiver. The $N_{\mathrm{f}}$ modes received are first split into two copies, one of which is sent to the off-diagonal multimode S-Comm receiver [Fig.~\ref{Fig:SCommOffDiagRX}], while the second copy is sent to a diagonal multimode S-Comm receiver. See Fig.~\ref{Fig:CombiningDiafOffDiag} for a conceptual schematic of this arrangement. We proceed to describe three approaches for encoding data streams into the diagonal elements of the coherence matrix $\mathbf{G}_{\mathrm{out}}(t)$.

\subsection{Using the standard definition of generalized Pauli matrices for Stokes multiplexing}

In general, $N_{\mathrm{f}}-1$ diagonal generalized Pauli matrices (or generalized Gell-Mann matrices) $\{\hat{\Lambda}_{N_{\mathrm{S}}+\ell}\}_{\ell=1}^{N_{\mathrm{f}}-1}$ are needed, whose standard definition is \cite{Kimura03PLA,Bertlmann08JPA}:
\begin{equation}
\hat{\Lambda}_{N_{\mathrm{S}}+\ell}=\sqrt{\frac{2}{\ell(\ell+1)}}\left(\sum_{j=1}^{\ell}|j\rangle\langle j|-\ell|\ell+1\rangle\langle\ell+1|\right),
\end{equation}
where $1\leq\ell\leq N_{\mathrm{f}}-1$. It is straightforward to verify that the properties of generalized Pauli matrices hold here.

When $N_{\mathrm{f}}=N_{\mathrm{S}}=2$, we recover the usual diagonal Pauli matrix, $\hat{\sigma}_{3}=\sqrt{\tfrac{2}{1\times2}}(|1\rangle\langle1|-|2\rangle\langle2|)=\left(\!\!\begin{array}{cc}1&0\\0&-1\end{array}\!\!\right)$. When $N_{\mathrm{f}}=3$ and $N_{\mathrm{S}}=6$, we have two diagonal matrices ($\ell=1,2$):
\begin{equation}
\hat{\Lambda}_{7}=\sqrt{\frac{2}{1\times2}}\left(|1\rangle\langle1|-|2\rangle\langle2|\right)=\left(\begin{array}{ccc}1&0&0\\0&-1&0\\0&0&0\end{array}\right),\nonumber
\end{equation}
\begin{equation}\label{eq:Lambda7Lambda8}
\hat{\Lambda}_{8}=\sqrt{\frac{2}{2\times3}}\left(|1\rangle\langle1|+|2\rangle\langle2|-2|3\rangle\langle3|\right)=\frac{1}{\sqrt{3}}\left(\!\!\begin{array}{ccc}1&0&0\\0&1&0\\0&0&-2\end{array}\!\!\!\right)\!.
\end{equation}
We write these diagonal matrices in contracted form as: $\hat{\Lambda}_{7}=\mathrm{diag}\{1,-1,0\}$ and $\hat{\Lambda}_{8}=\tfrac{1}{\sqrt{3}}\mathrm{diag}\{1,1,-2\}$, where $\mathrm{diag}\{\cdot\}$ refers to a diagonal matrix with the entries in the parentheses corresponding to its diagonal elements.

\begin{figure}[t!]
\centering
\includegraphics[width=8.8cm]{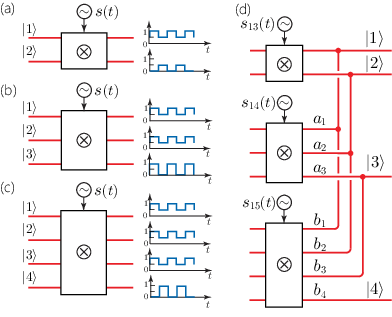}
\caption{(a) Design of coherence modulators for bimodes, (b) trimodes, (c) and 4-modes. We plot on the right the modulated signals associated with the modes. In each case, the total power in each bit remains constant. The bimode coherence modulator in (a) corresponds to $\hat{\sigma}_{3}$ when $N_{\mathrm{f}}=2$, $\hat{\Lambda}_{7}$ when $N_{\mathrm{f}}=3$, $\hat{\Lambda}_{13}$ when $N_{\mathrm{f}}=4$, $\hat{\Lambda}_{21}$ when $N_{\mathrm{f}}=5$, and $\hat{\Lambda}_{31}$ when $N_{\mathrm{f}}=6$ (in general, $\hat{\Lambda}_{N_{\mathrm{S}}+1}$ for $N_{\mathrm{f}}\geq2$). The trimode coherence modulator in (b) corresponds to $\hat{\Lambda}_{8}$ when $N_{\mathrm{f}}=3$, $\hat{\Lambda}_{14}$ when $N_{\mathrm{f}}=4$, $\hat{\Lambda}_{22}$ when $N_{\mathrm{f}}=5$, and $\hat{\Lambda}_{32}$ when $N_{\mathrm{f}}=6$ (in general, $\hat{\Lambda}_{N_{\mathrm{S}}+2}$ for $N_{\mathrm{f}}\geq3$). The 4-mode coherence modulator in (c) corresponds to $\hat{\Lambda}_{15}$ for $N_{\mathrm{f}}=4$, $\hat{\Lambda}_{23}$ when $N_{\mathrm{f}}=5$, and $\hat{\Lambda}_{33}$ when $N_{\mathrm{f}}=6$ (in general $\hat{\Lambda}_{N_{\mathrm{S}}+3}$ for $N_{\mathrm{f}}\geq4$). The output modes require appropriate amplitude scaling factors when combined (dictated by the normalization factors). (d) Stokes multiplexing of the modes from the coherence modulators associated with $\hat{\Lambda}_{13}$, $\hat{\Lambda}_{14}$, and $\hat{\Lambda}_{15}$ for $N_{\mathrm{f}}=4$; $(a_{1},a_{2},a_{3})$ and $(b_{1},b_{2},b_{3},b_{4})$ are amplitudes factors to accommodate the correct normalizations.}
\label{Fig:CoherenceModulators}
\end{figure}

The~3 diagonal Pauli matrices for $N_{\mathrm{f}}=4$ and $N_{\mathrm{S}}=12$ are:
\begin{eqnarray}\label{eq:Lambda13_14_15}
\hat{\Lambda}_{13}&=&\mathrm{diag}\{1,-1,0,0\},\nonumber\\
\hat{\Lambda}_{14}&=&\tfrac{1}{\sqrt{3}}\mathrm{diag}\{1,1,-2,0\},\nonumber\\
\hat{\Lambda}_{15}&=&\tfrac{1}{\sqrt{6}}\mathrm{diag}\{1,1,1,-3\}.
\end{eqnarray}
Finally, for future reference, the 4~diagonal Pauli matrices for $N_{\mathrm{f}}=5$ and $N_{\mathrm{S}}=20$ are:
\begin{eqnarray}
\hat{\Lambda}_{21}&=&\mathrm{diag}\{1,-1,0,0,0\},\nonumber\\
\hat{\Lambda}_{22}&=&\tfrac{1}{\sqrt{3}}\mathrm{diag}\{1,1,-2,0,0\},\nonumber\\
\hat{\Lambda}_{23}&=&\tfrac{1}{\sqrt{6}}\mathrm{diag}\{1,1,1,-3,0\},\nonumber\\
\hat{\Lambda}_{24}&=&\tfrac{1}{\sqrt{10}}\mathrm{diag}\{1,1,1,1,-4\};
\end{eqnarray}
and the~5 diagonal Pauli matrices for $N_{\mathrm{f}}=6$ and $N_{\mathrm{S}}=30$ are:
\begin{eqnarray}
\hat{\Lambda}_{31}&=&\mathrm{diag}\{1,-1,0,0,0,0\},\nonumber\\
\hat{\Lambda}_{32}&=&\tfrac{1}{\sqrt{3}}\mathrm{diag}\{1,1,-2,0,0,0\},\nonumber\\
\hat{\Lambda}_{33}&=&\tfrac{1}{\sqrt{6}}\mathrm{diag}\{1,1,1,-3,0,0\},\nonumber\\
\hat{\Lambda}_{34}&=&\tfrac{1}{\sqrt{10}}\mathrm{diag}\{1,1,1,1,-4,0\},\nonumber\\
\hat{\Lambda}_{34}&=&\tfrac{1}{\sqrt{10}}\mathrm{diag}\{1,1,1,1,1,-5\}.
\end{eqnarray}

A hierarchy of diagonal matrices can be discerned from the examples above. It is clear that for a field spanned by $N_{\mathrm{f}}+1$~modes, the~$N_{\mathrm{f}}$ associated diagonal Pauli matrices incorporate all the $N_{\mathrm{f}}-1$ diagonal Pauli matrices associated with the field spanned by $N_{\mathrm{f}}$~modes -- \textit{after} embedding them into the larger space (by adding a final row and column of zeros) -- in addition to one new diagonal matrix $\hat{\Lambda}_{N_{\mathrm{D}}}$ in which all the diagonal elements are non-zero. For example, fields spanned by $N_{\mathrm{f}}=3$ modes have two diagonal $3\times3$ Pauli matrices, $\hat{\Lambda}_{7}$ and $\hat{\Lambda}_{8}$ (Eq.~\ref{eq:Lambda7Lambda8}). Here $\hat{\Lambda}_{7}$ is simply $\hat{\sigma}_{3}$ for $N_{\mathrm{f}}=2$ embedded in the new higher-dimensional space. Moving up to fields spanned by $N_{\mathrm{f}}=4$ modes, we now have three diagonal $4\times4$ Pauli matrices; $\hat{\Lambda}_{13}$, $\hat{\Lambda}_{14}$, and $\hat{\Lambda}_{15}$ (Eq.~\ref{eq:Lambda13_14_15}). The diagonal $4\times4$ matrices $\hat{\Lambda}_{13}$ and $\hat{\Lambda}_{14}$ for $N_{\mathrm{f}}=4$ are nothing but the diagonal $3\times3$ matrices $\hat{\Lambda}_{7}$ and $\hat{\Lambda}_{8}$ for $N_{\mathrm{f}}=3$ after embedding them into the larger-dimensional space (the last row and column in $\hat{\Lambda}_{13}$ and $\hat{\Lambda}_{14}$ are all zeros). A new diagonal matrix $\hat{\Lambda}_{15}$ is added for $N_{\mathrm{f}}=4$ whose diagonal elements are all non-zero (Eq.~\ref{eq:Lambda13_14_15}). Taking the next step to $N_{\mathrm{f}}=5$, we now have 4~diagonal $5\times5$ matrices $\hat{\Lambda}_{21}$, $\hat{\Lambda}_{22}$, $\hat{\Lambda}_{23}$, and $\hat{\Lambda}_{24}$, the first three of which ($\hat{\Lambda}_{21}$, $\hat{\Lambda}_{22}$, and $\hat{\Lambda}_{23}$) are nothing but the $\hat{\Lambda}_{13}$, $\hat{\Lambda}_{14}$, and $\hat{\Lambda}_{15}$ associated with $N_{\mathrm{f}}=4$ after embedding them in the larger-dimensional space. A new diagonal matrix $\hat{\Lambda}_{24}$ is added whose diagonal elements are all non-zero. 

A challenge posed by this standard definition of diagonal Pauli matrices is that the corresponding coherence modulators are each structured differently: it receives a different number of uncorrelated modes, and new modulation profiles for the modes are required. For example, when $N_{\mathrm{f}}=3$, the coherence modulator associated with the diagonal Pauli matrix $\hat{\Lambda}_{7}$ coincides with that used for bimodes~\cite{Abouraddy26JOSAA1}; see Fig.~\ref{Fig:BimodalTX} and Fig.~\ref{Fig:CoherenceModulators}(a). However, the coherence modulator associated with $\hat{\Lambda}_{8}$ requires modulating three uncorrelated modes (all diagonal elements of $\hat{\Lambda}_{8}$ are non-zero; Eq.~\ref{eq:Lambda7Lambda8}). Consequently, this trimode coherence modulator requires three uncorrelated modes and a different modulation scheme implemented by the underlying conventional modulators as depicted in Fig.~\ref{Fig:CoherenceModulators}(b). A similar challenge occurs in $N_{\mathrm{f}}=4$, where the bimode and trimode field structures for $N_{\mathrm{f}}=3$ correspond to $\hat{\Lambda}_{13}$ and $\hat{\Lambda}_{14}$, respectively, whereas the field corresponding to $\hat{\Lambda}_{15}$ requires 4~uncorrelated modes fed into a new coherence modulator depicted in Fig.~\ref{Fig:CoherenceModulators}(c). 

After obtaining the coherence-modulated fields $\mathbf{D}_{N_{\mathrm{S}}+\ell}(t)$, the fields are then Stokes-multiplexed, just as is the case of the off-diagonal coherence-modulated bimodes with no need for the additional step of Stokes-encoding. However, because of the different normalization factors involved in the definition of the diagonal Pauli matrices, different amplitude scaling factors are needed for the bimodes, which can in principle be absorbed into the construction of the coherence modulators. An example of Stokes-multiplexing to the diagonal SPs is shown in Fig.~\ref{Fig:CoherenceModulators}(d) for $N_{\mathrm{f}}=4$.

\subsection{Alternative approach for Stokes multiplexing}

The standard definition of diagonal generalized Pauli matrices requires constructing coherence modulators for the diagonal SPs that differ from those used for the off-diagonal SPs (which are identical to the bimodal case). Of course, linearly independent superpositions of these diagonal Pauli matrices can be used as an alternative basis. One possibility is to make use of diagonal matrices of the form:
\begin{eqnarray}
\hat{\Lambda}_{N_{\mathrm{S}}+1}\!\!\!&=&\!\!\!\mathrm{diag}\{1,-1,0,0,0,0,\cdots\},\nonumber\\
\hat{\Lambda}_{N_{\mathrm{S}}+2}\!\!\!&=&\!\!\!\mathrm{diag}\{0,0,1,-1,0,0,\cdots\},\nonumber\\
\hat{\Lambda}_{N_{\mathrm{S}}+3}\!\!\!&=&\!\!\!\mathrm{diag}\{0,0,0,0,1,-1,\cdots\},
\end{eqnarray}
and so on. In other words, we embed $\hat{\sigma}_{3}$ in the $N_{\mathrm{f}}\times N_{\mathrm{f}}$ space and shift it along the diagonal. These matrices obviously satisfy the requirements for generalized Pauli matrices. This provides $N_{\mathrm{f}}/2$ matrices for even~$N_{\mathrm{f}}$ and $(N_{\mathrm{f}}-1)/2$ for odd~$N_{\mathrm{f}}$. The advantage of this approach is that its implementation makes use of the same bimodal coherence modulator utilized for the off-diagonal SPs. The disadvantage of course is that it does not exploit the full information-carrying capacity  of the field ($\sim N_{\mathrm{f}}/2$ rather than a possible $N_{\mathrm{f}}-1$ diagonally multiplexed data streams).

A hierarchy of diagonal matrices can be defined that add more diagonal data streams. For example, for $N_{\mathrm{f}}=4$ we have:
\begin{eqnarray}\label{eq:ModifiedN4}
\hat{\Lambda}_{13}'\!\!\!&=&\!\!\!\mathrm{diag}\{1,-1,0,0\},\nonumber\\
\hat{\Lambda}_{14}'\!\!\!&=&\!\!\!\mathrm{diag}\{0,0,1,-1\},\nonumber\\
\hat{\Lambda}_{15}'\!\!\!&=&\!\!\!\tfrac{1}{\sqrt{2}}\mathrm{diag}\{1,1,-1,-1\}.
\end{eqnarray}
Whereupon the structure of the required coherence modulators differ minimally from those of the off-diagonal SPs. This approach may also reduce the number of connections needed for Stokes multiplexing. In establishing the modified diagonal Pauli matrices for odd~$N_{\mathrm{f}}$, we simply take the diagonal matrices for the preceding number of even modes and then add a single matrix corresponding to the standard definition. Therefore, when $N_{\mathrm{f}}=3$ we add $\hat{\Lambda}_{8}=\tfrac{1}{\sqrt{3}}\mathrm{diag}\{1,1,-2\}$ after $\hat{\Lambda}_{7}=\mathrm{diag}\{1,-1,0\}$, which is simply $\hat{\sigma}_{3}$ embedded in the $3\times3$ space. When $N_{\mathrm{f}}=5$ we make use of the diagonal matrices for $N_{\mathrm{f}}=4$ in Eq.~\ref{eq:ModifiedN4}, and then we have:
\begin{eqnarray}
\hat{\Lambda}_{21}'\!\!\!&=&\!\!\!\mathrm{diag}\{1,-1,0,0,0\},\nonumber\\
\hat{\Lambda}_{22}'\!\!\!&=&\!\!\!\mathrm{diag}\{0,0,1,-1,0\},\nonumber\\
\hat{\Lambda}_{23}'\!\!\!&=&\!\!\!\tfrac{1}{\sqrt{2}}\mathrm{diag}\{1,1,-1,-1,0\},\nonumber\\
\hat{\Lambda}_{24}\!\!\!&=&\!\!\!\tfrac{1}{\sqrt{10}}\mathrm{diag}\{1,1,1,1,-4\}.
\end{eqnarray}
For $N_{\mathrm{f}}=6$ we have:
\begin{eqnarray}
\hat{\Lambda}_{31}'\!\!\!&=&\!\!\!\mathrm{diag}\{1,-1,0,0,0,0\},\nonumber\\
\hat{\Lambda}_{32}'\!\!\!&=&\!\!\!\tfrac{1}{\sqrt{3}}\mathrm{diag}\{1,1,-2,0,0,0\},\nonumber\\
\hat{\Lambda}_{33}'\!\!\!&=&\!\!\!\mathrm{diag}\{0,0,0,0,1,-1\},\nonumber\\
\hat{\Lambda}_{34}'\!\!\!&=&\!\!\!\tfrac{1}{\sqrt{3}}\mathrm{diag}\{0,0,0,1,1,-2\},\nonumber\\
\hat{\Lambda}_{35}'\!\!\!&=&\!\!\!\tfrac{1}{\sqrt{3}}\mathrm{diag}\{1,1,1,-1,-1,-1\},
\end{eqnarray}
which combine the results for $N_{\mathrm{f}}=2$ and $N_{\mathrm{f}}=3$. This strategy can be extended further for each value of $N_{\mathrm{f}}$, but more work is needed to find the optimal set of diagonal Pauli matrices for large $N_{\mathrm{f}}$ that requires minimal deviation from the off-diagonal coherence modulators for their implementation.

\begin{figure}[t!]
\centering
\includegraphics[width=8.8cm]{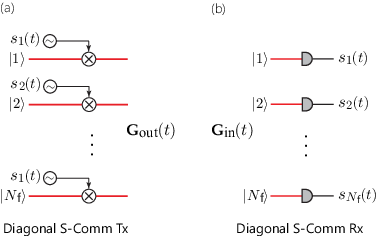}
\caption{(a) Schematic of the hybrid S-Comm transmitter for the diagonal data streams. Each data stream $s_{j}(t)$ modulates a single mode from $N_{\mathrm{f}}$ uncorrelated modes. (b) Diagonal S-Comm receiver for hybrid multiplexing. Each detected mode reveals a data stream.}
\label{Fig:}
\end{figure}

\subsection{Hybrid Stokes multiplexing}

Rather than encoding data streams in diagonal SPs, one may instead encode data streams directly in the power of the diagonal elements, so that $G_{\ell\ell}(t)=s_{N_{\mathrm{S}}+\ell}(t)$, $\ell=1,\cdots,N_{\mathrm{f}}$. This definition can be viewed as associated with generalized Pauli matrices of the form:
\begin{equation}
\hat{\Lambda}_{N_{\mathrm{S}}+\ell}=|\ell\rangle\langle\ell|,\;\;\ell=1,\cdots,N_{\mathrm{f}}.
\end{equation}
Rather than using coherence modulators, we rely on conventional modulators that modulate a single mode via a data stream, thus requiring only $N_{\mathrm{f}}$~modulators. This is a hybrid approach because the off-diagonal data streams make use of generalized SPs for their encoding, whereas the remaining data streams encoded in the diagonal elements do not make use of the corresponding SPs and instead rely on conventional modulation. Moreover, rather than $N_{\mathrm{f}}-1$ data streams encoded in the diagonal SPs, this hybrid approach provides encoding for $N_{\mathrm{f}}$ data streams. The total number of multiplexed data streams in both off-diagonal and diagonal elements of $\mathbf{G}$ is thus $N_{\mathrm{S}}+N_{\mathrm{f}}=N_{\mathrm{D}}+1=N_{\mathrm{f}}^{2}$. Therefore, bimodal fields with $N_{\mathrm{f}}=2$ can encode $N_{\mathrm{f}}^{2}=4$ data streams in the hybrid S-Comm scheme. The drawback of this approach is that the power per bit changes, in contrast to the approaches outlined above for encoding data streams in diagonal SPs where the power per bit remains constant. In this hybrid scheme, the coherence matrices for $N_{\mathrm{f}}=2$ and $N_{\mathrm{f}}=3$, for example, are:
\begin{equation}\label{eq:Hybrid2matrix}
\mathbf{G}_{\mathrm{out}}(t)\propto\left(\begin{array}{cc}
2+s_{3}(t)&s_{1}(t)-is_{2}(t)\\
s_{1}(t)+is_{2}(t)&2+s_{4}(t)
\end{array}\right),
\end{equation}
\begin{equation}\label{eq:Hybrid3matrix}
\mathbf{G}_{\mathrm{out}}(t)\propto\left(\begin{array}{ccc}
2+s_{7}(t)&s_{1}(t)-is_{2}(t)&s_{3}(t)-is_{4}(t)\\
s_{1}(t)+is_{2}(t)&2+s_{8}(t)&s_{5}(t)-is_{6}(t)\\
s_{3}+is_{4}(t)&s_{5}(t)+is_{6}(t)&2+s_{9}(t)
\end{array}\right),
\end{equation}
respectively.

\subsection{Diagonal Stokes demultiplexing}

In all three approaches to multiplexing data streams in the diagonal elements of $\mathbf{G}$, the $N_{\mathrm{f}}$~modes underpinning $\mathbf{G}_{\mathrm{in}}(t)$ at the S-Comm receiver are split into two copies, one of which is directed to the \textit{off-diagonal} S-Comm receiver, while the second is directed to a \textit{diagonal} S-Comm receiver [Fig.~\ref{Fig:CombiningDiafOffDiag}]. A different diagonal S-Comm receiver can be implemented for each of the three diagonal S-Comm transmitter schemes described above (based on the standard definition, the alternative definition, and the hybrid approach). However, all three diagonal multiplexing approaches can be \textit{demultiplexed} simply by detecting the modal weights, corresponding to the diagonal elements of the coherence matrix $\mathbf{G}_{\mathrm{in}}(t)$. In the case of hybrid Stokes multiplexing, the modal weights provide the data streams directly atop a constant background (Eq.~\ref{eq:Hybrid2matrix} and Eq.~\ref{eq:Hybrid3matrix}, for example). In the other two diagonal-SP multiplexing schemes, additional circuitry is needed to extract the data streams, from combinations of the modal weights. The alternative definition of diagonal Pauli matrices simplify the modal-weight combinations needed with respect to those associated with the conventional definition.


\section{Discussion}

\subsection{Comparison to mutual-coherence communications}

We briefly compare S-Comm with an alternative approach for dense optical communications proposed in Ref.~\cite{Nardi22OL} that also makes use of multimode partially coherent light, which we refer to as mutual-coherence communications (MC-Comm).

\subsubsection{Number of multiplexed data streams}

Whereas S-Comm multiplexes $N_{\mathrm{D}}=N_{\mathrm{f}}^{2}-1$ independent data streams in $N_{\mathrm{f}}$ modes, the MC-Comm scheme proposes multiplexing $\tfrac{1}{2}N_{\mathrm{S}}=\tfrac{1}{2}N_{\mathrm{f}}(N_{\mathrm{f}}-1)$ data streams in $N_{\mathrm{f}}$~modes (only a single data stream is encoded in each off-diagonal element). The number of multiplexed data streams in both S-Comm and MC-Comm grow quadratically with $N_{\mathrm{f}}$ but at different rates. For $N_{\mathrm{f}}=2,3,4,5$, the number of multiplexed data streams in MC-Comm is $\tfrac{1}{2}N_{\mathrm{S}}=1,3,6,10$, so that an improvement only occurs when $N_{\mathrm{f}}\geq4$, in contrast to $N_{\mathrm{D}}=3,8,15,24$ in S-Comm, so that there is always an enhancement. Moreover, MC-Comm is based on a version of the coherence matrix normalized so that all the diagonal elements are equal, thus precluding their utilization for encoding data streams.

\subsubsection{Construction of the transmitter}

As emphasized in Ref.~\cite{Abouraddy26JOSAA1}, it is crucial that: (1) any multiplexer or demultiplexer in high-speed optical communications be a \textit{fixed} device with no need to change its structure from bit to bit; and (2) the demultiplexer is a multiplexer operated in reverse. Both of these are features of conventional multiplexing systems for the spectral, polarization, and spatial DoFs. The S-Comm multiplexing systems described here for $N_{\mathrm{f}}>2$ and in Ref.~\cite{Abouraddy26JOSAA1} for $N_{\mathrm{f}}=2$ also satisfy these desiderata. The MC-Comm transmitter proposed in Ref.~\cite{Nardi22OL} takes as input $N_{\mathrm{f}}$ generic uncorrelated modes; i.e., the initial coherence matrix is $\mathbf{G}=\tfrac{1}{N_{\mathrm{f}}}\hat{\mathbb{I}}_{N_{\mathrm{f}}}$.
Rather than modulate the input optical signals before traversing a fixed optical system as in S-Comm, the $\tfrac{1}{2}N_{\mathrm{S}}$ data streams in MC-Comm modulate the elements of a linear non-unitary system, such that the data streams are encoded in the off-diagonal elements of the mutual coherence matrix \cite{Nardi22OL}. Consequently, the MC-Comm multiplexer is \textit{not} a fixed device. Instead, the multipexer structure is required to change from bit to bit; e.g., by modulating phase shifters. Although this scheme can in principle multiplex $\tfrac{1}{2}N_{\mathrm{S}}$ data streams in $N_{\mathrm{f}}$~modes, it would be challenging to operate at high data rates.

\subsubsection{Construction of the receiver}

The MC-Comm receiver takes copies of pairs of modes from the $N_{\mathrm{f}}$~received modes and extracts the magnitude of an off-diagonal element of $\mathbf{G}$ by registering the visibility of an interference pattern. This requires scanning a relative phase between each pair of modes \textit{during a bit}. This is challenging in high-speed communications. Moreover, the multiplexing and demultiplexing systems are completely distinct rather than being the inverse of each other. In contrast, the S-Comm receiver is a fixed device that is basically the inverse of the S-Comm transmitter.

\subsection{On-chip implementation of S-Comm}

Crucially, all the subsystems in the S-Comm transmitter and receiver can be implemented in integrated photonic circuits as we demonstrated in our recent work regarding on-chip manipulation of multimode partially coherent light \cite{Hashemi26APLPhotonTwoMode,Hashemi26arxiv4Modes,Hashemi26LPR3modes,Hashemi26OL,Mor26arxiv}. Specifically, we have already realized all the Stokes-encoding and Stokes-decoding unitaries ($\hat{U}_{j}$ and $\hat{V}_{j}$, respectively) in arrays of on-chip Mach-Zehnder interferometers \cite{Bogaerts20Nature,Capmany20Book}. The standard optical modulators involved in the construction of the coherence modulators can be fiber-based and implemented off-chip, with the coherence-modulated bimodes $\mathbf{D}_{j}(t)$ then fed to an optical chip, or on-chip high-speed modulators can be utilized instead. The detectors at the final stage in the S-Comm receiver may also be implemented on-chip or off-chip.

\section{Conclusion}

In conclusion, we have provided explicit designs for constructing a Stokes communications (S-Comm) system that exploits multimode ($N_{\mathrm{f}}>2$) partially coherent light for dense optical communications. The construction of the transmitter and receiver that multiplex and demultiplex the data streams in the off-diagonal and diagonal Stokes parameters (SPs) required different approaches. Whereas the system for off-diagonal SPs makes use of the bimodal system as a unit cell, the system for diagonal SPs requires designing new components (specifically, multimode coherence modulators). These results pave the way to a new form of optical communications that exploits the elements of the coherence matrix associated with multimode partially coherent light as information carriers. This provides novel forms of secure and dense optical communications and may impact optical signal processing in general.

\textbf{Acknowledgments.} This work was funded by the U.S. Office of Naval Research (ONR) under contract N00014-20-1-2789.

\bibliography{diffraction}

\end{document}